\documentclass[12pt]{article}
\usepackage{amsmath,amssymb}
\newcommand{\be}{\begin{equation}}
\newcommand{\ee}{\end{equation}}
\newcommand{\bea}{\begin{eqnarray}}
\newcommand{\eea}{\end{eqnarray}}
\newcommand{\nn}{\nonumber \\}
\newcommand{\p}[1]{(\ref{#1})}
\newcommand{\lb}{\label}

\begin{document}
\begin{titlepage}

\begin{center}
{\Large\bf Supersymmetrizing Landau Models}
\vspace{1.5cm}

{\large\bf
Evgeny Ivanov}
\vspace{1cm}

{\it Bogoliubov Laboratory of Theoretical Physics,
JINR, \\
141980, Dubna, Moscow Region, Russia}\\
{\tt eivanov@theor.jinr.ru}\\[8pt]
\end{center}
\vspace{2cm}

\begin{abstract}
\noindent
This is an overview of recent progress in constructing and studying 
superextensions of the Landau problem of a quantum particle on a plane in the uniform 
magnetic field, as well as of its Haldane's $S^2$ generalization 
({\tt hep-th/0311159, 0404108, 0510019, 0612300}). The main attention is paid to 
the planar super Landau models which are invariant under the inhomogeneous supergroup $ISU(1|1)$, 
a contraction of the supergroup $SU(2|1)$, and provide minimal superextensions of 
the original Landau model. Their common notable feature is the 
presence of a hidden dynamical worldline ${\cal N}=2$ supersymmetry. It exists at 
the classical and quantum levels and is revealed most naturally while passing to 
the new invariant inner products in the space of quantum states in order to make the 
norms of all states positive. For one of the planar models, the superplane Landau model, 
we present an off-shell worldline superfield formulation in which the ${\cal N}=2$ 
supersymmetry gets manifest.

\end{abstract}
\vspace{6cm}

\begin{center}
{\it Talk given at the International Workshop on Classical and Quantum Integrable
Systems, Bogoliubov Laboratory of Theoretical Physics, JINR, Dubna, January 22--25, 2007}
\end{center}

\end{titlepage}

\setcounter{page}{1}
\section{Introduction}
The original Landau model \cite{Landau1930} describes a charged particle moving on a plane under the influence of a
uniform magnetic field orthogonal to the plane. A spherical version of the Landau model (Haldane's model \cite{Haldane1983}) describes a charged particle on a sphere
$S^2 \sim SU(2)/U(1)$ in the background of Dirac monopole placed in the centrum. This sort of models has plenty of applications, in particular, in the Quantum Hall Effect (QHE) \cite{QHE}.

Supersymmetric extensions of the Landau models we shall deal with in this talk are models of non-relativistic superparticles moving on supergroup manifolds. The study of such models can help to reveal possible manifestations of supersymmetry in various versions of QHE (including the so called spin-QHE) and other realistic condensed matter systems. From the matematical point of view,
Landau model and its superextensions bear close relations to non-commutative (super)geometry: 
in these models, after quantization, there arises a one-to-one correspondence between the Lowest Landau Levels (LLL) 
and non(anti)commutative (super)manifolds which are spanned by the relevant position operators.

The Landau problems on the $(2|2)$-dimensional supersphere $SU(2|1)/U(1|1)$ and the $(2|4)$-dimensional
superflag $SU(2|1)/[U(1)\times U(1)]$, minimal superextensions of the $S^2$ Haldane model, were considered  
in \cite{IMT1,IMT2} (see also \cite{Hasebe2}).
In order to better understand the salient features of the super Landau models, it was instructive
to study the planar limits of such supermanifold extensions of the original Landau models.
The planar models should follow from their curved supermanifold prototypes in the limit of the large 
$S^2$ radius (contraction). Such models were constructed and studied in \cite{IMT3,CIMT}, as well as
in \cite{Hasebe}. The model obtained in this way from the $SU(2|1)/U(1|1)$ one is called 
``the superplane Landau model'', while the model following from the $SU(2|1)/[U(1)\times U(1)]$ one 
is called ``the planar superflag Landau model''.

This contribution is a review of the super Landau models, with the main focus on their planar 
versions and the appearance of hidden worldline ${\cal N}=2$ supersymmetry in them, 
which seems to be a common feature of the planar models \cite{Hasebe,CIMT}. For the superplane Landau model, 
as the essentially new result, we develop the ${\cal N}=2$ superfield formalism making manifest  
its worldline ${\cal N}=2$ supersymmetry. We also address the problem of the presence of negative norm 
states in the quantum planar super Landau models \cite{IMT3} and demonstrate that in all cases this difficulty can be evaded by introducing a non-trivial metric on the space of states \cite{CIMT} along the lines of refs. \cite{Bend,Bi}.

\section{Preliminaries: bosonic Landau models}

\subsection{Bosonic Landau  models as d=1 WZW sigma models}

\noindent{\it 1. Definitions.} The standard planar Landau  model is described by the following Lagrangian
\be
L_b = |\dot z|^2 - i\kappa\left(\dot z\bar z - \dot{\bar z} z\right)= |\dot z|^2 + 
\left(A_z \dot z + A_{\bar z}\dot{\bar z}\right). \label{Llagr}
\ee
Here $z(t), \bar z(t)$ parametrize 2-dimensional Euclidean plane, $2\kappa$ is the value of a uniform  external magnetic field,
\be
A_z = -i\kappa \bar z,   A_{\bar z} = i\kappa z, \quad
\partial_{\bar z}A_z - \partial_{z}A_{\bar z} = - 2i\kappa\,.
\ee
The second term in \p{Llagr} is the simplest example of Wess-Zumino term.

The corresponding canonical Hamiltonian is represented as
\be
H =  \frac{1}{2}\left( a^{\dagger}a + aa^{\dagger}\right) = a^{\dagger}a + \kappa\,,
\ee
where
\be
a= i(\partial_{\bar z} + \kappa z), \quad a^\dagger = 
i(\partial_{z} - \kappa \bar z), \quad [a, a^\dagger] = 2\kappa\,. \label{Ham1}
\ee
It commutes with the following operators, which so define the invariances of the theory (see e.g. 
\cite{ElPo}):
\bea
&& P_z = -i(\partial_{z} + \kappa \bar z), P_{\bar z} =-i(\partial_{\bar z} - \kappa z),
 [P_z,  P_{\bar z}] = 2\kappa,\; F_b = z\partial_{z} - \bar z\partial_{\bar z}\,, \nn
&& [H, P_z] = [H, P_{\bar z}] = [H, F_b] =0\,.
\eea
The operators $P_z, P_{\bar z}$ and $F_b$ generate, respectively, target $2D$ ``magnetic translations''
and rotations. These operators are quantum versions of the Noether charges corresponding to 
the target space translations, $z'= z + a,
\bar z' = \bar z + \bar a$, and $U(1)$ rotations, $z'= e^{i\alpha}z, \bar z' = e^{-i\alpha}\bar z$.
\vspace{0.3cm}

\noindent{\it 2. Wavefunctions.}
The wavefunction of the lowest Landau level (LLL), $H \Psi_{(0)} = \kappa \Psi_{(0)}$, is defined by the following equation
\be
a \Psi_{(0)}(z, \bar z)  = 0 \quad \Leftrightarrow \quad (\partial_{\bar z} + \kappa z)\Psi_{(0)} = 0 \, \rightarrow \,
\Psi_{(0)} = e^{-\kappa |z|^2}\psi_{(0)}(z),
\ee
i.e. it is reduced to a holomorphic function.

The wavefunction corresponding to the $n$-th LL is constructed as
\be
\Psi_{(n)}(z, \bar z) = [i(\partial_{z} - \kappa \bar z)]^n  e^{-\kappa |z|^2}\psi_{(n)}(z),
\; H\Psi_{(n)} = \kappa (2n + 1)\Psi_{(n)}\,,
\ee
i.e. it is also expressed through a holomorphic function. Each LL is infinitely degenerate due to 
the $(P_z, P_{\bar z})$ invariance. At each level, the wavefunctions form infinite-dimensional 
irreps of this group with the basis $z^m, m>0$ \cite{ElPo}.

The invariant norm is defined by
\be
||\Psi_{(n)}||^2 = \int dz d\bar z\, \overline{\Psi}_{(n)}(z, \bar z)\Psi_{(n)}(z, \bar z)
\sim \int dz d\bar z\, e^{-2\kappa |z|^2}\overline{\psi}_{(n)}(\bar z)\psi_{(n)}(z) < \infty\,.
\ee
This integral is finite for any monomial $\psi_{(n)}(z) \sim z^m\,$.
\vspace{0.3cm}

\noindent{\it 3. Planar Landau model as a WZW model.}
Let us treat $2 \kappa $ in $[P_z,  P_{\bar z}] = 2\kappa $ as an independent generator
(``central charge'') and construct a nonlinear realization of this non-abelian magnetic translation group 
in a coset over the one-dimensional subgroup generated by $2 \kappa $. Choosing the exponential parametrization
for the relevant coset representatives, we have 
\bea
&& g(z, \bar z) = e^{i(zP_z + \bar z P_{\bar z})}, \quad g^{-1}d g = i\omega_z P_z + i\omega_{\bar z} P_{\bar z} + i\omega_\kappa 2\kappa\,, \nn 
&& \omega_z= dz, \;\; \omega_{\bar z} = d\bar z, \;\; \omega_\kappa = \frac{1}{2i}\left(z d\bar z - \bar z d z \right).
\eea
We observe that WZ term in $L_b$ is none other than the Cartan 1-form associated with $2\kappa$.
In this geometric setting, the creation and annihilation operators $a$ and $a^\dagger $ appear as the covariant derivatives
\be
\nabla_z =  \partial_{z} - \kappa \bar z, \quad \nabla_{\bar z} = \partial_{\bar z} + \kappa z\,,
\ee
while the LLL wavefunction is defined by the covariant Cauchy-Riemann condition
\be
\nabla_{\bar z}\Psi_{(0)} = 0, \; 
\Psi_{(n)}'(z', \bar z') = e^{-\kappa(a\bar z -\bar a z)} \Psi_{(n)}(z, \bar z)\,, \, z' = z + a\,, 
\bar z' = \bar z + \bar a\,.
\ee
\vspace{0.3cm}

\noindent{\it 2. Generalization to $S^2$.}
An $S^2$ analog of the planar Landau Lagrangian $L_b$ (i.e. the Haldane's model Lagrangian) reads
\be
\tilde{L}_b = \frac{1}{(1 + r^{2} |z|^2)^2}|\dot z|^2 - i s \frac{1}{1 +r^{2}|z|^2}\left(\dot z\bar z
- \dot{\bar z} z\right).
\ee
The first term is the $d=1$ pullback of the invariant interval on $S^2$, the second term is the $d=1$ WZ term 
on the coset $SU(2)/U(1)$, $r$ is the ``inverse'' radius of $S^2$.

Upon quantization, the unitary wavefunctions at each level are finite-dimensional $SU(2)$ irreps,
$s, s+1/2, s+ 1, \ldots $ being their ``spins''. So, each LL is finitely degenerated (this finite degeneration is the important distinction from the planar case, and it is of course related to the fact that $SU(2)$ is compact group while its contraction, the magnetic translation group, is non-compact). The LLL wavefunction is determined by the covariant analyticity condition on $S^2$
\be
\nabla_{\bar z}\Psi_{(0)} = 0, \quad \nabla_{\bar z} = (1 + r^2|z|^2)\partial_{\bar z}
+ \mbox{$U(1)$ connection}\,.
\ee
This wavefunction can be shown to reduce to the holomorphic function of $z$ encompassing the spin $s$ $SU(2)$ irrep.
The next LL functions are constructed similarly to the planar case, and the corresponding 
wavefunctions are expressed through holomorphic functions representing $SU(2)$ irreps with spins $s+1/2, s+ 1, \ldots $. 
The limit $r\rightarrow 0$ takes us back to the planar Landau model.

\subsection{Lowest Landau Level and space-time non-commutativity}
Let us come back to the case of the planar Landau model. 
The LLL and 1st LL energy gap is $E_1 - E_0 = \kappa$, and so it goes to $\infty$ with $\kappa$ growing. In this limit 
only the LLL survives and it is described by the pure WZ term. The analyticity condition $\nabla_{\bar z}\Psi_{(0)}= 0$ 
is recovered as the quantum version of the corresponding phase space constraint.
The standard position operator $Z= z$ does not commute with this constraint and should be modified:
\be
Z = z - \frac{1}{\kappa}\partial_{\bar z} = \frac{1}{i\kappa}P_{\bar z}\,, \; \bar Z =\frac{i}{\kappa}P_{z}\,, \; [\bar Z, Z] = \frac{2}{\kappa}.
\ee
As a result, $Z, \bar Z$ parametrize a non-commutative plane.

In the $S^2$ case the situation is similar: in the LLL limit only WZ term survives and the
corresponding position operators commuting with the Hamiltonian constraints parametrize  
non-commutative (``fuzzy'') version of $S^2$ \cite{Madore1992}.

\setcounter{equation}{0}
\section{Supersymmetrizing Landau model: toy example}
Super-Landau models are quantum-mechanical models for a charged particle on a homogeneous superspace,
such that their ``bosonic'' truncation yields either Landau's original model for a charged
particle moving on a plane under the influence of a uniform magnetic field,
or Haldane's spherical version of it.

\subsection{Worldline supersymmetry vs target space supersymmetry}
One way to obtain a supersymmetric extension of the Landau models is through the 
worldline supersymmetrization:
\bea
&& t \; \Rightarrow \; (t, \theta, \bar\theta), \qquad z, \bar z  \; \Rightarrow \; {\cal Z}(t, \theta, \bar\theta), \;
\bar{{\cal Z}}(t, \theta, \bar\theta)\,, \nn 
&& z, \bar z \;\Rightarrow \; (z, \bar z , \psi, \bar\psi, \ldots ) - \mbox{worldline supermultiplet}.
\eea
Here ${\cal Z}$ are some worldline superfields. The resulting models provide a version of supersymmetric quantum 
mechanics \cite{Witten} (see e.g. \cite{APa} ). In this scheme of supersymmetrization, the fermionic fields have no 
immediate geometric interpretation, as distinct from the bosonic fieldfs $z, \bar z$ which, as we saw, 
are some coset parameters.

Another way is to extend the Landau models by introducing a kind of target space supersymmetry.
The simplest option is to enlarge the group of magnetic translations to a supergroup involving the 
``magnetic supertranslation'' generators $\Pi_\zeta, \Pi_{\bar\zeta}$ and a fermionic analog of the 
$U(1)$ generator $F_b$: 
\bea
&& \mbox{group manifold}: \; (z, \bar z)  \;\; \Rightarrow \;\; \mbox{supergroup manifold}: \;
(z, \bar z, \zeta, \bar\zeta)\,, \nn 
&& (P_z, P_{\bar z}, F_b, \kappa)\;\;\Rightarrow \;\;
(P_z, P_{\bar z}, \Pi_\zeta, \Pi_{\bar\zeta}, F_b, F_f, \kappa, \ldots ), \nn 
&& \Pi_\zeta = \partial_{\zeta} + \kappa \bar \zeta\,,\;\; \Pi_{\bar\zeta}
= \partial_{\bar\zeta} + \kappa \zeta\,,\;\;
F_f = \zeta\partial_{\zeta} -\bar \zeta\partial_{\bar\zeta}\,, \;\;
\{\Pi_\zeta, \Pi_{\bar\zeta}\} = 2\kappa\,. \label{MagSup}
\eea
In this case the additional fermionic fields have a clear geometrical meaning: 
they are Grassmann coordinates extending 2-dimensional plane to a $(2|2)$-dimensional superplane.

It is remarkable that two planar super Landau models constructed by using the second approach (see Sections 4 and 5) 
reveal, as a ``gift'', an emergent hidden worldline ${\cal N}=2$ supersymmetry.
\subsection{Fermionic ``Landau model''}

Before turning to the discussion of super Landau models it is instructive to look at the 
toy ``fermionic Landau model'', with the bosonic $2D$ coordinates $z, \bar z$ being 
replaced by the fermionic ones $\zeta, \bar\zeta$. 

The corresponding Lagrangian and quantum Hamiltonian read
\bea
&& L_f=\dot\zeta\dot{\bar\zeta} - i\kappa\left(\dot\zeta \bar\zeta
+ \dot{\bar\zeta}\zeta\right),\quad H_{f}=\frac{1}{2}\left[\alpha, \alpha^\dagger\right] =
-\alpha^{\dagger}\alpha-\kappa\,, \nn
&& \alpha = \partial_{\bar\zeta}-\kappa\zeta,\, \alpha^\dagger = \partial_{\zeta}-\kappa\bar\zeta,\,
\{\alpha,\, \alpha^\dagger\} = -2\kappa\,.
\eea
The invariances are given by the generators $\Pi_\zeta, \Pi_{\bar\zeta}$ and $F_f$ defined 
in \p{MagSup}:
\be
[H_f, \Pi_\zeta] = [H_f,\Pi_{\bar\zeta}] = [H_f,F_f] = 0\,.
\ee
The quantum ``Hilbert space'' of this model consists of the ground state and single excited state:
\bea
&& \psi^{(0)} = e^{-\kappa\zeta\bar\zeta}\, \psi_0\left(\zeta\right), \;
\psi^{(1)}= e^{\kappa\zeta\bar\zeta}\, \psi_1\left(\bar\zeta\right),\;  \alpha \psi^{(0)} 
= \alpha^\dagger \psi^{(1)} = 0, \nn
&& \psi_0 =A_0 + \zeta\, B_0 \, ,\qquad \psi_1 = A_1 +  \bar\zeta\, B_1\,.
\eea
The pairs  $(A_0\,, B_0)\,$, $(A_1\,, B_1)$ form two irreps of the magnetic supertranslation group, with 
energies $-\kappa$ and $\kappa$.

On this simplest example we encounter a problem which is also revealed in the fermionic extensions of the 
Landau model considered in the next Sections. It is the appearance of ghosts and the necessity to ``exorcize'' them.

With the natural supertranslation invariant choice of the inner product, 
\bea
&& <\phi \big| \psi > =\int d\zeta d\bar{\zeta}~\overline
{\phi\left(  \zeta,{\bar{\zeta}}\right)  }\psi\left(  \zeta,\bar{\zeta}
{}\right), \, \label{fInn}
\eea
one finds:
\bea
&& <\psi^{(0)}\big|\psi^{(1)}> = 0, \nn
&& < \psi^{(0)}\big|\psi^{(0)}>  =2\kappa\bar{A}_{0}A_{0}
+\bar{B}_{0}B_{0}\,,\,
<\psi^{(1)}\big|\psi^{(1)}> =-2\kappa\bar{A}_{1}A_{1}
-\bar{B}_{1}B_{1}\,.
\eea
We see that the states $A_1, B_1$ have negative norm, i.e. they are ghosts \footnote{The appearance of ghosts in $d=1$ 
supersymmetric models with the second order kinetic term for fermions was noted in \cite{VoPa}.}. The presence of ghosts
is as usual an unpleasant property since it can give rise to breaking of unitarity \footnote{See, however, \cite{Andr}.}. However, in the considered 
case one can cure this difficulty by introducing a non-trivial metric
on the ``Hilbert space'': 
\be
<<\phi \big| \psi >> := <G \phi \big| \psi >\,, \quad G\left(\psi^{(0)}+\psi^{(1)}\right) = 
\psi^{(0)}-\psi^{(1)}\,, \;G= - \kappa^{-1} H_f\,.
\ee

The characteristic features of this procedure can be summarized as follows.
\begin{itemize}
\item Symmetry generators $\Pi_\zeta, \Pi_{\bar\zeta}$ and $F_f$ commute with the metric $G$, 
so the new inner product remains invariant.
However, the hermitian conjugation properties of the operators which do not commute with $G$, change, e.g.
$$
\alpha^{\ddagger}=-\alpha^{\dagger}\, \Rightarrow \,
H_{f}=\alpha^{\ddagger}\alpha-\kappa\,, \;\{ \alpha,\alpha^{\ddagger}\}  = 2\kappa\,.
$$
\item Under the new conjugation the momentum canonically conjugate
to a coordinate is also the coordinate's hermitian conjugate:
$$
\zeta^{\ddagger}={\frac{1}{\kappa}}\partial_{\zeta}\ ,\; \left(  \bar
{\zeta}\right)  ^{\ddagger}={\frac{1}{\kappa}}\partial_{\bar{\zeta}}\ ,\;
\left(\partial_\zeta\right)^\ddagger = \kappa\zeta\, ,\;
\left(\partial_{\bar\zeta}\right)^\ddagger = \kappa\bar\zeta\,.
$$
\end{itemize}

So much for the toy model. Now let us turn to our main subject.

\setcounter{equation}{0}
\section{Superplane Landau model}

\subsection{Lagrangian, Hamiltonian and symmetries}

The superplane Landau model is a hybrid of the bosonic and fermionic Landau models. It is described by 
the following Lagrangian
\be
L= L_f+L_b = |\dot z|^2 + \dot\zeta\dot{\bar\zeta}-
i\kappa\left(\dot z\bar z - \dot{\bar z} z + \dot\zeta \bar\zeta
+ \dot{\bar\zeta}\zeta\right). \label{spLact}
\ee
The corresponding quantum Hamiltonian reads
\be 
H = a^{\dagger}a - \alpha^\dagger \alpha  =
\partial_{\bar\zeta}\partial_{\zeta}- \partial_{z}\partial_{\bar z} +
\kappa\left(  \bar z \partial_{\bar z} + \bar\zeta\partial_{\bar\zeta} - z
\partial_{z} - \zeta\partial_{\zeta}\right)  + \kappa^{2}\left(  z\bar z +
\zeta\bar\zeta\right).
\ee

The set of invariances, besides those generated by $P_z, P_{\bar z}$ and $\Pi_\zeta, \Pi_{\bar\zeta}$, 
involves the new ones, with the generators
\bea
&& Q=z\partial_{\zeta}-\bar{\zeta}\partial_{\bar{z}}\,,\; Q^{\dagger}=\bar
{z}\partial_{\bar{\zeta}}+\zeta\partial_{z}\,, \; C= F_b + F_f = z\partial_{z}+\zeta\partial_{\zeta}
-\bar{z}\partial_{\bar{z}}
-\bar{\zeta}\partial_{\bar{\zeta}}\,, \label{isu11gen} \\
&& [H, Q] = [H, Q^{\dagger }] = [H, C] = 0\,.
\eea
These generators form the superalgebra $ISU(1|1)$, a contraction of the superalgebra $SU(2|1)\,$,
\be
\{Q,Q^{\dagger}\}=C\,,\; [C, Q] = [C, Q^\dagger] =0\,, \; [Q,P_z]=i\Pi_\zeta\, ,\; \{Q^\dagger,\Pi_\zeta\}=iP_z\,.
\ee

\subsection{States and degeneracies}
The LLL wavefunction $\psi^{(0)}$ is defined to vanish under both the bosonic and fermionic 
annihilation operators $a$ and $\alpha$
\bea
&& \left( \partial_{\bar{z}}+\kappa z\right)\psi^{(0)} = \left(  \partial_{\bar{\zeta}}
-\kappa\zeta\right)\psi^{(0)} =0 \,\Rightarrow \nn
&& \psi^{(0)} = e^{-\kappa K_2}\, \psi^{(0)}_{an}(z,\zeta)\,, \,
K_2 = |z|^2 + \zeta{\bar \zeta}\,, \quad H \psi^{(0)} = 0,
\eea
and so it has the extra two-fold degeneracy, $\psi^{(0)}_{an}(z,\zeta) = A^{(0)}(z) + \zeta B^{(0)}(z)\,$.

The $N$-th level Hilbert space is spanned by a wavefunction
\bea
&& \psi^{(N)} \sim  \left(a^\dagger\right)^N e^{-\kappa K_2} \psi^{(N)}_+ \left(z,\zeta\right)
+ \left(a^\dagger\right)^{N-1}\alpha^\dagger e^{-\kappa K_2}\psi^{(N)}_-\left(z,\zeta\right), \nn
&& H\psi^{(N)} = 2\kappa N \psi^{(N)},
\eea
with $\psi^{(N)}_\pm(z,\zeta) = A^{(N)}_\pm (z) + \zeta B^{(N)}_\pm (z) $. So, each LL 
for $N>0$  reveals a 4-fold degeneracy.

The natural  $ISU(1|1)$-invariant inner product,
\be
<\phi \big|\psi> =\int \!d\mu\
\overline{\phi\left( z,\bar z;\zeta, \bar \zeta\right)}\,
\psi\left(z,\bar z; \zeta,\bar\zeta\right),\; d\mu = dzd\bar{z}d\zeta d\bar{\zeta}\,,
\ee
leads to negative norms for some states, like in the fermionic toy model.
However, all norms can be made positive by introducing the same metric on the Hilbert space
\be
G=- \kappa^{-1}H_f = \frac{1}{\kappa}\left[  \partial_{\zeta}\partial_{\bar\zeta} +\kappa^{2}
\bar\zeta\zeta+ \kappa\left(  \zeta\partial_{\zeta}- \bar\zeta\partial
_{\bar\zeta}\right)  \right].
\ee
The metric $G$ commutes with all symmetry generators, except for $Q$, $Q^\dagger$.
Hence, with respect to the new inner product $<G \phi \big| \psi >$, the conjugate of $Q$ is different 
from $Q^\dagger$. This new conjugate $Q^{\ddagger}$ is easily calculated: 
\be
Q^{\ddagger}=Q^{\dagger}- \frac{i}{\kappa}\,  S\,, \, S=i\left(  \partial_{z}\partial_{\bar\zeta} +
\kappa^{2} \bar z \zeta-\kappa\bar z \partial_{\bar\zeta} - \kappa
\zeta\partial_{z}\right).
\ee
Since both operators $Q^\dagger$ and $Q^{\ddagger}$ commute with the full Hamiltonian 
$H$ (because $[H, G] = 0$), their difference $S$ also commutes, $[H, S]= [H, S^\ddagger] = 0$,
and hence defines a new (hidden) symmetry of the model.

\subsection{Hidden ${\cal N}=2$ worldline supersymmetry}

The operators $S, S^\ddagger$ can be written as 
\be
S= a^{\dagger}\alpha\,, \;S^\ddagger = a\alpha^\ddagger
\ee
and easily checked to generate a worldline ${\cal N}=2$ supersymmetry:
\bea
[H, S] = [H, S^\ddagger] = 0\,, \quad \{S,S^\ddagger \} = 2\kappa H\,, \quad 
\{S,S\}=0=\{S^\ddagger,S^\ddagger\}\,.
\eea
In other words, ${(2k)}^{-1/2} S\,, \, {(2k)}^{-1/2}S^\ddagger$ and $H$ form ${\cal N}=2$, $d=1$ Poincar\'e superalgebra.

The LLL ground state is annihilated by $S, S^\ddagger$
\be
S\psi^{(0)} = S^\ddagger \psi^{(0)} = 0,
\ee
and so this state is the ${\cal N}=2$ supersymmetry singlet. Hence, ${\cal N}=2$ supersymmetry is unbroken, 
and all higher LL wavefunctions form irreps of it. Each state comprises two irreducible
$ISU(1|1)$ irreps, which explains the 4-fold degeneracy of LL with $N>0$.
It is interesting that there exists the following finite-dimensional  model-independent analog of the Sugawara 
representation for this worldline supersymmetry in terms of the $ISU(1|1)$ charges
\be
S=2i\kappa Q^{\ddagger}+P\Pi^{\ddagger}\ ,\, S^{\ddagger}
=-2i\kappa Q+P^{\ddagger}\Pi\,, \,H = P^{\ddagger}P + \Pi^{\ddagger}\Pi- 2\kappa C,
\ee
i.e. $S, S^\ddagger, H$ belongs to the enveloping algebra of $ISU(1|1)$.

Classically, the hidden wordline supersymmetry is realized as the following transformations
of $z, \zeta$ and their conjugates
\be
\delta z = \epsilon \dot\zeta  \, ,\qquad \delta \zeta = - \dot z \bar\epsilon\,.\label{spLtran}
\ee
They have the correct on-shell closure on time derivative, taking into account
the equations of motion $\ddot{z}= 2i\kappa \dot z,
\ddot{\zeta} = 2i\kappa \dot \zeta$. The ${\cal N}=2$ supersymmetry exists only at $\kappa \neq 0$, so it should be 
viewed as a sort of dynamical symmetry.

In Section 6 we show how to reproduce this realization from a manifestly ${\cal N}=2$ supersymmetric
off-shell superfield approach.

\setcounter{equation}{0}
\section{Planar superflag model}
Here we consider salient features of one more $ISU(1|1)$ invariant model extending the Landau model: 
the planar superflag Landau model \cite{IMT2}.

\subsection{Classical and quantum structure of the model}
\noindent{\it 1. Definitions.} The superflag Landau model \cite{IMT2} describes a charged particle on the coset
superspace $SU(2|1)/[U(1)\times U(1)]$. One of the two WZ terms
associated with the $U(1)\times U(1)$ group is the Lorentz coupling to
a uniform magnetic field of strength $2\kappa\,$. The second WZ term
is purely ``fermionic'', with the constant coefficient $M$.

In the planar limit, when $SU(2|1)$ contracts into $ISU(1|1)$ and
$S^2$ into the Euclidean  2-plane, the superflag action of \cite{IMT2} becomes \cite{IMT3}
\bea
L = (1+\bar\xi\xi)|\dot z|^2 + (\bar\xi \dot{\bar z} \dot \zeta
- \xi \dot z \dot{\bar\zeta}) + \bar\xi\xi \dot\zeta \dot{\bar\zeta}
- \ i\kappa (\dot z \bar z - \dot{\bar z} z+ \dot\zeta\bar\zeta
+ \dot{\bar\zeta} \zeta )
+iM (\bar\xi\dot\xi + \xi\dot{\bar\xi})\,.\label{LplSF}
\eea
The basic difference from the superplane model is the presence of the extra fermionic variable
$\xi(t), \bar\xi(t)$. It can be interpreted as the Nambu-Goldstone variable associated with
the $ISU(1|1)$ generators $Q, Q^\dagger$. Due to these additional variables one can construct 
the second WZ term and simultaneously avoid the appearance of the unconventional second-order kinetic term 
for $\zeta, \bar\zeta$. Despite these attractive features, there are still negative norms in the quantum 
theory, provided one uses the natural definition of the inner product.
\vspace{0.3cm}

\noindent{\it 2. Quantization.}
There are phase space constraints in the theory (due to the 1-st order fermionic terms).
Solving them, one finds that the generic wavefunction has the following structure
\be
\Psi = K_1^M e^{-\kappa K_2} \Psi_{ch}\left(z,\bar z_{sh}, \zeta,\xi\right),\;K_1 = 1+ \bar\xi\xi\,, \;
\bar z_{sh} = \bar z - \xi\bar\zeta\,.
\ee
The Hamiltonian operator, while acting on these ``physical'' wavefunctions, can be written as
\bea
&& H= \hat a^\dagger \hat a\,,\quad [\hat a,\hat a^\dagger] = 2\kappa\,, \nn
&& \hat a = i\sqrt{K_1}\left(\partial_{\bar z} + \kappa\, z_{sh}
- \bar\xi\partial_{\bar\zeta}\right), \,
\hat a^\dagger = i\sqrt{K_1}\left(\partial_z
- \kappa\, \bar z_{sh}- \xi\partial_\zeta \right).
\eea
At the Landau level $N$ the physical chiral wavefunction has the special structure: 
it is expressed through an {\it analytic} function of $(z,\zeta,\xi)$ as
\be
\Psi_{ch}^{(N)} = \tilde\nabla_z^N \Psi_{an}^{(N)}\left(z,\zeta,\xi\right), \;
\tilde\nabla_z = \partial_z -2\kappa \bar z_{sh} - \xi\partial_\zeta\,, \;H \Psi_{ch}^{(N)}
= 2\kappa N\Psi_{ch}^{(N)}\,.
\ee

The $ISU(1|1)$-invariant  inner product is naturally defined by
\be
<\Phi \big| \Psi> = \int \!d\mu \int\! d\xi d\bar\xi\ \overline{\Phi}\, \Psi \
= \int \! d\mu\,  e^{-2\kappa K_2}\! \int \! d\xi d\bar\xi\ K_1^{2M}\,
\overline{\Phi_{ch}} \, \Psi_{ch}\, ,
\ee
where $d\mu = dz d\bar z d\zeta d\bar\zeta $ is the superplane model integration measure.
Expanding
\be
\Psi_{an}^{(N)} = A^{(N)}(z) + \zeta B^{(N)}(z) + \xi C^{(N)}(z) + \zeta \xi B^{(N)}(z),
\ee
one finally finds
\be
||\Psi||^2 \propto \int dz d\bar z e^{-2\kappa |z|^2}[(2M-N)(2\kappa A^\dagger A + B^\dagger B)
+ 2\kappa C^\dagger C + D^\dagger D]\,.
\ee
So for $N > 2M >0$ and $M< 0$ there are negative norms. At $N = 2M$ there are zero norms.

Like in the previous cases, one can redefine the inner product by introducing a non-trivial
metric operator on the Hilbert space. On the analytic functions it is represented as
\be
G_{an} = [\xi,\partial_\xi] = -1 + 2\xi\partial_\xi\,.
\ee
After this redefinition, the norm (under the $z, \bar z$ integral) changes to
\be
\propto [(N -2M)(2\kappa A^\dagger A + B^\dagger B) + 2\kappa C^\dagger C + D^\dagger D]\,.
\ee
Now all states have a positive norm when $M{<}0$. This remains true for $M{=}0$ except
that half of the $N{=}0$ states have zero norm.
The (super)space of physical states can be naturally defined as the quotient by the subspace of zero-norm states.
Thus, zero-norm states do not contribute to the physical spectrum. Then 
it follows that the $M{=}0$ planar superflag model has precisely the same spectrum,
including degeneracies, as the superplane model, and is therefore equivalent to it.
At $M>0$ there still remain negative norms when $N<2M$ and we are led to use the old ``naive''
norm in this range of parameters.

\subsection{Hidden ${\cal N}=2$ supersymmetry}
As in the superplane model, passing to the new norm alters the hermitian conjugation properties of the
$ISU(1|1)$ supercharge $Q$ and, as a result, new conserved supercharges naturally appear. While acting on
the analytic wavzefunctions, these supercharges are given by the expressions
\be
S_{an} = 2i\kappa\xi \left(  2M-N_{an}\right),\,
 S_{an}^{\ddagger}= 2i\kappa\partial_\xi,\,\{S_{an},S_{an}^\ddagger\} = 2\kappa (H_{an} -4\kappa M).
\label{sflN2}
\ee
This quantum worldline supersymmetry exists for $M\leq 0$ since it is impossible to achieve
a positive definiteness of the anticommutator in \p{sflN2} at
$M>0$ in the full range of parameters and at every LL.

For $M=0$ the ground state ($N=0$) is annihilated by both $S$ and $S^\ddagger$,
modulo zero norm states. So in this case the worldline supersymmetry is unbroken, 
the ground state is the supersymmetry singlet and it has only double degeneracy due to
the $ISU(1|1)$ invariance. At $N>0$ the wavefunctions form non-trivial supersymmetry multiplets, each consisting
of two $ISU(1|1)$ irreps, whence the 4-fold degeneracy arises. The $M=0$ superflag is equivalent
to the superplane model. For $M<0$ there are no supersymmetry invariant ground state, so ${\cal N}=2$ supersymmetry
is spontaneously broken in this case.

As in the superplane case, the ${\cal N}=2$ supersymmetry possesses an on-shell realization
on the original $d=1$ field variables
\be
\delta z = -\epsilon\xi\left(\dot z + \bar\xi \dot\zeta\right)\, , \;
\delta\zeta =-\left[\left(1+\bar\xi\xi\right)\dot z + \bar\xi\dot\zeta\right]\bar\epsilon\, ,\;
\delta\xi = -2i\kappa\, \bar\epsilon\,
\ee
(the realization on $\bar z, \bar \zeta, \bar\xi$ is obtained by complex conjugation).
These transformations  close on time derivative for each field, after taking
into account the equations of motion. In particular, the correct on-shell closure on $\xi$ is achieved 
due to the property that $\dot\xi = 0$ on shell. The closure is proportional
to $\kappa$, so the ${\cal N}=2$ supersymmetry exists only at $\kappa \neq 0$ and therefore, as in the superplane 
Landau model,  is of the dynamical character. It would be of obvious interest to recover this realization 
and the action \p{LplSF} from some off-shell ${\cal N}=2$ superfield formalism, as we do this in the next Section 
for the case of the superplane model.

\setcounter{equation}{0}
\section{Worldline ${\cal N}=2$ supersymmetry made manifest}
Here we reformulate the superplane model 
of Sect. 4 within the worldline ${\cal N}=2$ superfield approach.

\subsection{Superfield action of the superplane model}

\noindent{\it 1. Definitions}. The basic objects are two ${\cal N}=2,\, d=1$ chiral bosonic and fermionic
superfields $\Phi$ and $\Psi$ of the same dimension. We use the following
conventions.

The real ${\cal N}=2,\, d=1$ superspace is parametrized as:
\be
(\tau, \theta, \bar\theta)\,. \lb{1}
\ee
The left and right chiral superspaces are defined by
\be
(t_L, \theta), \quad (t_R, \bar\theta), \quad t_L = \tau -i\theta\bar\theta, \; t_R
= \tau +i\theta\bar\theta = t_L + 2i\theta\bar\theta\,. \lb{2}
\ee
It will be convenient to work in the left (chiral) basis, so for brevity we will use
the notation $t_L \equiv t,\; t_R = t + 2i\theta\bar\theta$. In this basis, two
${\cal N}=2$ covariant derivatives are defined by
\be
\bar D = -\frac{\partial}{\partial \bar\theta}\,, \; D =\frac{\partial}{\partial \theta}
- 2i \bar\theta \partial_t\,, \; \{D, \bar D\} = 2i\partial_t\,, \; D^2 = \bar D^2 = 0\,. \lb{3}
\ee

The chiral superfields $\Phi$ and $\Psi$ obey the
basis-independent conditions
\be \bar D \Phi = \bar D\Psi = 0
\lb{Chir}
\ee
and in the left-chiral basis have the following
component field contents
\be \Phi (t, \theta) = z(t) + \theta
\chi(t)\,, \;\; \Psi(t, \theta) = \psi(t) + \theta h(t)\,,
\lb{4}
\ee
where the complex fields $z(t), h(t)$ are bosonic and
$\chi(t), \psi(t)$ are fermionic. The conjugated superfields, in
the same left-chiral basis, have the following $\theta$
-expansion
\be \bar \Phi = \bar z -\bar\theta \bar\chi + 2i
\theta\bar\theta \dot{\bar z}\,, \; \bar\Psi = \bar \psi +
\bar\theta \bar h  + 2i\theta\bar\theta \dot{\bar\psi}\,. \lb{5}
\ee
Also, we shall need the component structure of the following
superfields
\bea
D\Phi &=& \chi -2i \bar\theta \dot{z} + 2i
\theta\bar\theta \dot\chi\,, \; \bar D\bar\Phi = (D\Phi)^\dagger
= \bar\chi + 2i\theta \dot{\bar z}\,, \nn D\Psi &=& h
-2i\bar\theta \dot\psi + 2i\theta\bar\theta \dot h\,, \; \bar
D\bar\Psi = -(D\Psi)^\dagger = -\bar h  + 2i\theta
\dot{\bar\psi}\,. \lb{6}
\eea
The Berezin integral is normalized
as
\be \int d^2\theta (\theta\bar\theta) = 1\,.
\ee
\vspace{0.3cm}

\noindent{\it 2. Superfield action}. The superfield action yielding in components the superplane
model action \p{spLact} (up to a renormalization factor) is as follows
\be
S = - \int dtd^2\theta \left\{\Phi\bar\Phi + \Psi\bar\Psi + \rho\left[\Phi D\Psi -
\bar\Phi \bar D\bar\Psi\right]\right\} \equiv \int dtd^2\theta \left\{{\cal L}_1 +
{\cal L}_2\right\}\,.\lb{7}
\ee
Here $\rho$ is a real parameter. After doing the Berezin integral, we find
\bea
&&{\cal L}_1 \;\Rightarrow \; -2i\left(z\dot{\bar z} + \psi\dot{\bar\psi}\right)
-\left(\chi\bar\chi + h\bar h \right)\,, \nn
&&{\cal L}_2 \;\Rightarrow \; -2i\rho\left(z \dot h + \chi \dot\psi + \dot{\bar z} \bar h + 
\bar\chi \dot{\bar \psi}\right). \lb{8}
\eea
The fields $h$ and $\chi$ are auxiliary and they can be eliminated by their equations
of motion
\be
\chi = 2i\rho\, \dot{\bar\psi}\,, \quad h = -2i \rho\, \dot{\bar z}\,. \lb{9}
\ee
Upon substituting this into the sum ${\cal L} \equiv {\cal L}_1 + {\cal L}_2$, the latter 
becomes
\be
{\cal L} \;\Rightarrow \; -2i\left(z\dot{\bar z} + \psi\dot{\bar\psi}\right) +
4\rho^2 \left(\dot z \dot{\bar z} + \dot{\bar\psi}\dot\psi \right). \lb{10}
\ee
After redefining
\be
\bar\psi = \zeta\,, \; \psi = \bar\zeta\,, \quad 4 \rho^2 \equiv \frac{1}{\kappa}\,,\lb{RL}
\ee
and integrating by parts, the Lagrangian \p{10} takes the form
\be
{\cal L} = -i\left(z \dot{\bar z} - \bar z \dot z + \zeta\dot{\bar\zeta} - \dot\zeta\bar\zeta\right)
+ \frac{1}{\kappa}\left(\dot z \dot{\bar z} + \dot\zeta\dot{\bar\zeta} \right), \lb{11}
\ee
that is reduced to the superplane model Lagrangian \p{spLact} after reversing the time variable, $t \rightarrow -t$, and inserting the overall renormalization factor $-\kappa$ in front of the action.

\subsection{Symmetries}

\noindent{\it 1. Supersymmetry.} By construction, the superfield action \p{7} is manifestly ${\cal N}=2$
supersymmetric. The ${\cal N}=2$ transformations of the component fields are defined by 
\be
\delta \Phi = -\left[\epsilon Q - \bar\epsilon \bar Q\right]\Phi\,, \quad
\delta \Psi = -\left[\epsilon Q - \bar\epsilon \bar Q\right]\Psi\,, \lb{12}
\ee
where, in the left-chiral basis,
\be
Q = \frac{\partial}{\partial \theta}\,, \quad \bar Q = - \frac{\partial}{\partial \bar\theta}
- 2i\theta \partial_t\,, \quad \{Q, \bar Q \} = -2i\partial_t = 2 P_0\,. \lb{13}
\ee
It follows from \p{12}, \p{13} that off shell
\be
\delta z = - \epsilon \chi\,, \quad \delta \chi = 2i \bar\epsilon \dot z\,, \quad
\delta \psi = -\epsilon h\,, \quad \delta h = 2i \bar\epsilon \dot\psi\,. \lb{14}
\ee
With the on-shell values \p{9} for the auxiliary fields
and with taking into account the relabelling \p{RL}, these transformations become
\be
\delta z = - \frac{i}{\sqrt{k}}\, \epsilon \dot{\zeta}\,, \quad \delta \zeta = -  
\frac{i}{\sqrt{k}}\,\bar\epsilon \dot z\,.
\ee
This is basically the same transformation law as \p{spLtran} (up to rescaling of $\epsilon, \bar\epsilon $).
\vspace{0.3cm}

\noindent{\it 2. Two  $ISU(1|1)$ symmetries.} 
The superfield $ISU(1|1)$ transformations look a little bit more
complicated. The main constraint on them is that they should be
consistent with the chirality conditions \p{Chir}. While the target
supertranslations act as constant shifts of $\Phi $ and $\Psi $,
\be
\delta_b \Phi = b\,, \quad \delta_\beta \Psi = \nu\,, \lb{STr}
\ee
where $b$ and $\nu$ are complex constant parameters, bosonic
and fermionic, the off-shell odd $SU(1|1)$ transformations involve
explicit $\theta$ and $\bar\theta$:
\bea
&& \delta \Phi = \bar
D\left(\omega \bar\theta \bar\Psi -  \frac{1}{2\,\sqrt{k}}\,\bar\omega \theta
D\Phi\right), \; \delta \Psi = \bar D\left(\omega \bar\theta
\bar\Phi - \frac{1}{2\,\sqrt{k}}\, \bar\omega \theta D\Psi\right), \nn && \delta
\bar\Phi = D\left(\bar\omega \theta \Psi - \frac{1}{2\,\sqrt{k}}\, \omega \bar\theta
\bar D\bar \Phi\right), \; \delta \bar\Psi = -D\left(\bar\omega
\theta \Phi +  \frac{1}{2\sqrt{k}}\,\omega \bar\theta \bar D\bar\Psi\right), \lb{su11}
\eea
where $\omega$, $\bar\omega$ are the corresponding Grassmann
transformation parameters. They are consistent with the chirality and anti-chirality
of the superfield variations due to the presence of the projectors $\bar D$ and $D$
in their r.h.s. In components, these transformations yield
\be
\delta z = \omega \bar\psi\,, \; \delta \chi =  \frac{i}{\sqrt{k}}\,\bar\omega \dot z\,, \;
\delta \psi = \omega \bar z\,, \; \delta h =  \frac{i}{\sqrt{k}}\,\bar\omega \dot\psi \lb{su11comp}
\ee
(and c.c.). These close on $C$ transformations which are realized by
\be
\delta_c z = i\alpha z\,, \; \delta_c \psi = -i\alpha \psi\,, \;
\delta_c \chi = \frac{i}{\sqrt{k}}\, \alpha \dot{\bar\psi}\,, \; \delta_c h =
\frac{i}{\sqrt{k}}\, \alpha \dot{\bar z} \lb{Ccomp}
\ee
(and c.c.), where $\alpha$ is the transformation parameter. The on-shell expressions
for the auxiliary fields \p{7} are compatible with the transformations \p{su11comp}, \p{Ccomp}.
The transformations \p{su11comp} and \p{Ccomp} just match with the form of the $ISU(1|1)$ 
generators \p{isu11gen}.

Due to the presence of explicit $\theta$ s in \p{su11}, these $SU(1|1)$ transformations do not commute with
the worldline ${\cal N}=2$ supersymmetry. 
It is amusing that there exists another type of ``internal'' odd transformations which
by construction commute with the ${\cal N}=2$ supersymmetry
\be
\delta \Phi = \lambda \left(\Psi +  \frac{1}{2\,\sqrt{k}}\,\bar D\bar\Phi\right), \;
\delta \Psi = \bar\lambda \left(\Phi - \frac{1}{2\,\sqrt{k}}\, \bar D\bar\Psi\right), \lb{New}
\ee
where $\lambda$ is the corresponding Grassmann parameter.
For the component fields, \p{New} imply the following transformations:
\bea
&&\delta z = \lambda\left(\psi + \frac{1}{2\,\sqrt{k}}\, \bar\chi \right)\; \Rightarrow \;
\lambda\left(\psi -\frac{i}{2\,k}\, \dot\psi \right), \lb{A} \\
&& \delta\psi =  \bar\lambda\left(z + \frac{1}{2\,\sqrt{k}}\, \bar h \right)\; \Rightarrow \;
\bar\lambda\left(z + \frac{i}{2\,k}\, \dot z \right),\lb{B} \\
&&\delta\chi =  -\lambda\left(h + \frac{i}{\sqrt{k}}\,\dot{\bar z}\right)\; \Rightarrow \; 0\,, \nn
&&\delta h =  -\bar\lambda\left(\chi - \frac{i}{\sqrt{k}}\,\dot{\bar\psi}\right)\; \Rightarrow \; 0\,, \lb{C}
\eea
where the arrow indicates the on-shell variations (with the auxiliary fields being eliminated
by \p{9}). It is straightforward to check invariance of \p{10} (or \p{11}) under \p{A}, \p{B}.
The transformations \p{New} also generate $SU(1|1)$ supergroup which is different from the
``standard'' one \p{su11} - \p{Ccomp}. Computing the Lie bracket of two transformations
\p{C} we can find the explicit superfield realization of the relevant $\tilde{C}$ transformation:
\be
\delta_{\tilde{c}} \Phi = i\beta \left(\Phi + \frac{i}{2\,k}\,\dot \Phi\right), \quad
\delta_{\tilde{c}} \Psi = i\beta \left(\Psi + \frac{1}{\sqrt{k}}\,\bar D\bar\Phi + 
\frac{i}{2\,k}\, \dot \Psi\right).\lb{Cnew}
\ee
These transformations commute with \p{New}, as expected. Together with the target
supertranslations \p{STr}, the $SU(1|1)$ transformations \p{New} and \p{Cnew} form
the other $ISU(1|1)$ supergroup. This supergroup does not commute with the ``standard'' one (though
both have the common target supertranslations sector): the closure contains some new transformations
which seemingly generate a kind of infinite-dimensional target space super-symplectic diffeomorphism
group (the presence of hidden symmetries of this sort in the on-shell bosonic action \p{10} was mentioned
in \cite{IMT3}).

It still remains to construct ${\cal N}=2$ superfield formulation of the planar superflag model.
It should operate with the same superfields $\Phi$ and $\Psi$ plus the new Goldstone spinor
chiral superfield $\Omega = \xi(t) + \theta g(t)$. The first-order form of the component action \p{LplSF} given 
in \cite{IMT3} could be the appropriate starting point for such a construction. 
An interesting further problem is to construct the models like the superflag one, starting from 
the ${\cal N}=2$ superfield formalism. Also it seems that the above construction can be extended 
to yield generalizations of the superplane and planar superflag models with hidden 
${\cal N}=4$ worldline supersymmetry (and hopefully with higher ${\cal N}$ ones).

\section{Summary and outlook}

Let me briefly summarize the contents of the talk.

\begin{itemize}
  \item Self-consistent superextensions of the bosonic Landau model
  can be constructed \cite{IMT1} - \cite{Hasebe} and they are WZW type sigma models on
    the appropriate graded extensions of the ``magnetic translation'' group underlying
    the original Landau model.
  \item Despite the appearance of bad negative norms for the ``natural'' choice 
of the invariant inner product,
  this can be circumvented by redefining the inner product, in line with the general 
methods of refs. \cite{Bend}, \cite{Bi}.
  \item An intriguing  common feature of the planar super Landau models is the appearance
  of the hidden dynamical ${\cal N}=2$ supersymmetry \cite{Hasebe,CIMT}.
  Does this property extend to the supersphere and superflag Landau models? Is it possible to recover
    it in all cases from a worldline  superfield formalism, as in the superplane model? 
What about higher ${\cal N}$? It would be desirable to get answers to these questions.
  \item As for possible physical applications, it would be tempting to understand 
in full what kind of phenomena is described by supersymmetric versions of QHE 
and what the physical meaning of the additional fermionic variables is in this context.
  \end{itemize}

\bigskip
\noindent {\bf Acknowledgements.} I would like to thank the Organizers of the CQIS-07 Workshop for 
inviting me to give this talk. Most of the results reported here were obtained in collaboration 
with Thomas Curtright, Luca Mezincescu and Paul K. Townsend to whom I express my deep gratitude. 
I acknowledge a support from the RFBR grant 06-02-16684, the grant INTAS-05-7928 and a grant of 
the Heisenberg-Landau program. I thank Laboratoire de Physique, UMR5672 of CNRS and  ENS-Lyon, 
for the kind hospitality at the final stage of this work and Francois Delduc for valuable remarks.

\end{document}